\documentclass[a4paper]{article}

\usepackage{INTERSPEECH2022}
\usepackage{graphics}

\title{JETS: Jointly Training FastSpeech2 and HiFi-GAN for End to End Text to Speech}
\name{Dan Lim, Sunghee Jung, Eesung Kim}
\address{
  Kakao Enterprise Corporation, Seongnam, Republic of Korea
  }
\email{\{satoshi.2020, ronda.jung, chris.ekim\}@kakaoenterprise.com}

\begin{document}

\maketitle
\begin{abstract}
In neural text-to-speech (TTS), two-stage system or a cascade of separately learned models have shown synthesis quality close to human speech. For example, FastSpeech2 transforms an input text to a mel-spectrogram and then HiFi-GAN generates a raw waveform from a mel-spectogram where they are called an acoustic feature generator and a neural vocoder respectively. However, their training pipeline is somewhat cumbersome in that it requires a fine-tuning and an accurate speech-text alignment for optimal performance. In this work, we present end-to-end text-to-speech (E2E-TTS) model which has a simplified training pipeline and outperforms a cascade of separately learned models. Specifically, our proposed model is jointly trained FastSpeech2 and HiFi-GAN with an alignment module. Since there is no acoustic feature mismatch between training and inference, it does not requires fine-tuning. Furthermore, we remove dependency on an external speech-text alignment tool by adopting an alignment learning objective in our joint training framework. Experiments on LJSpeech corpus shows that the proposed model outperforms publicly available, state-of-the-art implementations of ESPNet2-TTS on subjective evaluation (MOS) and some objective evaluations.

\end{abstract}

\section{Introduction}
Text-to-speech (TTS) based on the neural network has significantly improved synthesized speech quality over the past years. Generally, a task of neural TTS is divided into more manageable sub-tasks using an acoustic feature generator and a neural vocoder. In this two-stage system, an acoustic feature generator generates an acoustic feature from an input text first and then a neural vocoder synthesizes a raw waveform from an acoustic feature. Those models are trained separately and then joined for inference. An acoustic feature generator can be autoregressive and attention-based for implicit speech-text alignments \cite{shen2018natural}, \cite{li2019neural} or it can be non-autoregressive for efficient parallel inference and duration informed for robustness on synthesis error \cite{ren2019fastspeech}, \cite{yu2020durianx}, \cite{DBLP:conf/iclr/0006H0QZZL21}. There are lots of research on neural vocoder as well and some of the famous, widely used include \cite{oord2016wavenet}, \cite{kalchbrenner2018efficient}, normalizing flow based one \cite{prenger2019waveglow} and generative adversarial network (GAN) based ones \cite{kumar2019melgan}, \cite{yamamoto2020parallel}, \cite{kong2020hifi}, \cite{jang21_interspeech}.

Although the two-stage system is the dominant approach for TTS, training two models separately may result in degradation of synthesis quality due to an acoustic feature mismatch. Note that a neural vocoder takes the ground-truth acoustic features for training and the predicted ones from an acoustic feature generator for inference. For optimal performance, we can further train a pre-trained neural vocoder with predicted acoustic features, which is called fine-tuning \cite{jang21_interspeech}, \cite{pmlr-v139-kim21f}. Or we can train a neural vocoder with predicted acoustic feature from the beginning \cite{shen2018natural}. However, both methods make the training pipeline somewhat complicated in that the former requires additional training steps and the latter requires completion of training of an acoustic feature generator prior to vocoder training stage.

On the other hand, end-to-end text-to-speech (E2E-TTS) \cite{DBLP:conf/iclr/0006H0QZZL21}, \cite{pmlr-v139-kim21f}, \cite{donahue2021endtoend}, \cite{miao2021efficienttts}, \cite{weiss2021wave} is a recent research trend in which a speech waveform is directly generated from an input text in a single stage without distinction between an acoustic feature generator and a neural vocoder. Although there is no intermediate conversion to human-designed acoustic features such as mel-spectrogram, it has shown comparable performance to the two-stage TTS systems. Since E2E-TTS doesn't have a problem of an acoustic feature mismatch, it usually doesn't require fine-tuning or sequential training. Moreover, some works \cite{pmlr-v139-kim21f}, \cite{donahue2021endtoend} further simplify the training pipeline by incorporating an alignment learning module so that the model can be trained without dependency on external speech-text alignments tools.

In this work, we propose E2E-TTS with a simplified training pipeline and high-quality speech synthesis. Our work is similar to \cite{hayashi2021espnet2} in that joint training of an acoustic feature generator and a neural vocoder is researched and the experiments are based on the ESPNet2 toolkit. However, our proposed model directly synthesizes raw waveform from an input text without an intermediate mel-spectrogram. Moreover, we incorporate an alignment learning objective so that the proposed model can be trained in single-stage training without dependency on external alignments models. The contributions of our work can be summarized as follows.

\begin{itemize}
\item We make the E2E-TTS model by jointly training an acoustic feature generator and a neural vocoder, which are FastSpeech2 and HiFi-GAN respectively. It does not require pre-training or fine-tuning and it synthesizes high-quality speech without an intermediate mel-spectrogram.
\item We leverage an alignment learning framework \cite{9747707} to obtain token duration on the fly during the training. Thus the training of our proposed model does not require external speech-text alignments models.
\item The proposed model outperforms state-of-the-art implementations of ESPNet2-TTS \cite{hayashi2021espnet2} on both subjective and objective evaluations.
\end{itemize}

\section{Related work}
There are several E2E-TTS research that directly generates speech waveform from an input text. For examples, FastSpeech2s \cite{DBLP:conf/iclr/0006H0QZZL21} is similar to our work in that it uses FastSpeech2 and GAN-based vocoder; Parallel WaveGAN \cite{yamamoto2020parallel}. However, it requires an auxiliary mel-spectrogram decoder and a preparation of speech-text alignments to train the model. Although LiteTTS \cite{nguyen21e_interspeech} also combines an acoustic feature generator with HiFi-GAN, it still depends on external alignments models and focuses more on lightweight structures for on-device uses. 

On the other hand, EATS \cite{donahue2021endtoend} integrates alignment learning into its adversarial training framework and it improves alignment learning stability by employing soft dynamic time warping to spectrogram prediction loss. VITS \cite{pmlr-v139-kim21f} also learns alignments during the training in the process of maximizing the likelihood of data and it improves expressiveness by utilizing variational inference and normalizing flow in an adversarial training framework. In EFTS-Wav \cite{miao2021efficienttts}, they adopt MelGAN and devise a novel monotonic alignment strategy with mel-spectrogram decoder for alignment learning. Wave-Tacotron \cite{weiss2021wave} adopts an attention-based Tacotron \cite{shen2018natural} with the normalizing flow and it is optimized to simply maximize the likelihood of the training data.

In \cite{hayashi2021espnet2}, joint training of an acoustic feature generator and a neural vocoder was conducted and it proved its effectiveness at solving the problem of acoustic feature mismatch by showing significant improvement compared to the separately learned model. However, the performance of the jointly trained model could not match that of a separately learned, fine-tuned model.

\section{Model description}
The proposed model is E2E-TTS which is jointly trained FastSpeech2 and HiFi-GAN with an alignment module. In this section, we describe each component in order.

\begin{figure}[t]
\centering
\includegraphics[width=\columnwidth]{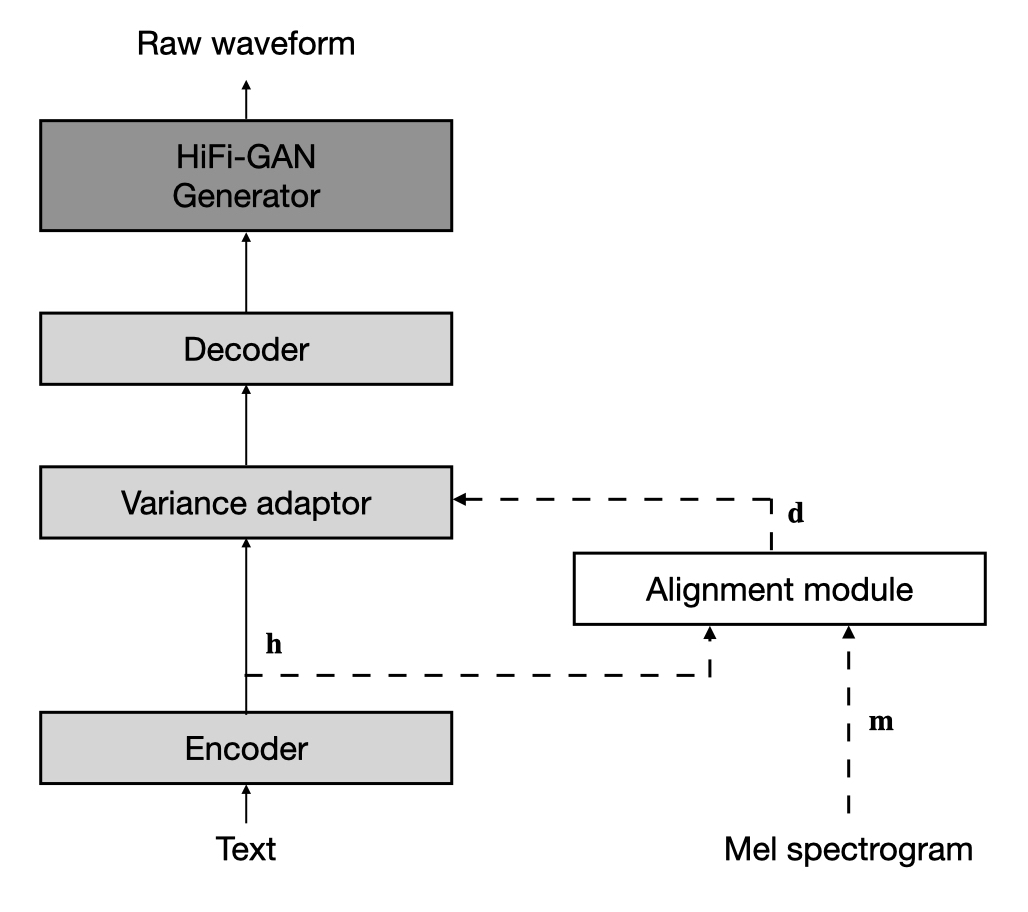}
\caption{An architecture of proposed model (discriminators are omitted for brevity)}
\label{fig:jets}
\end{figure}

\begin{figure}[t]
\centering
\includegraphics[width=0.7\columnwidth]{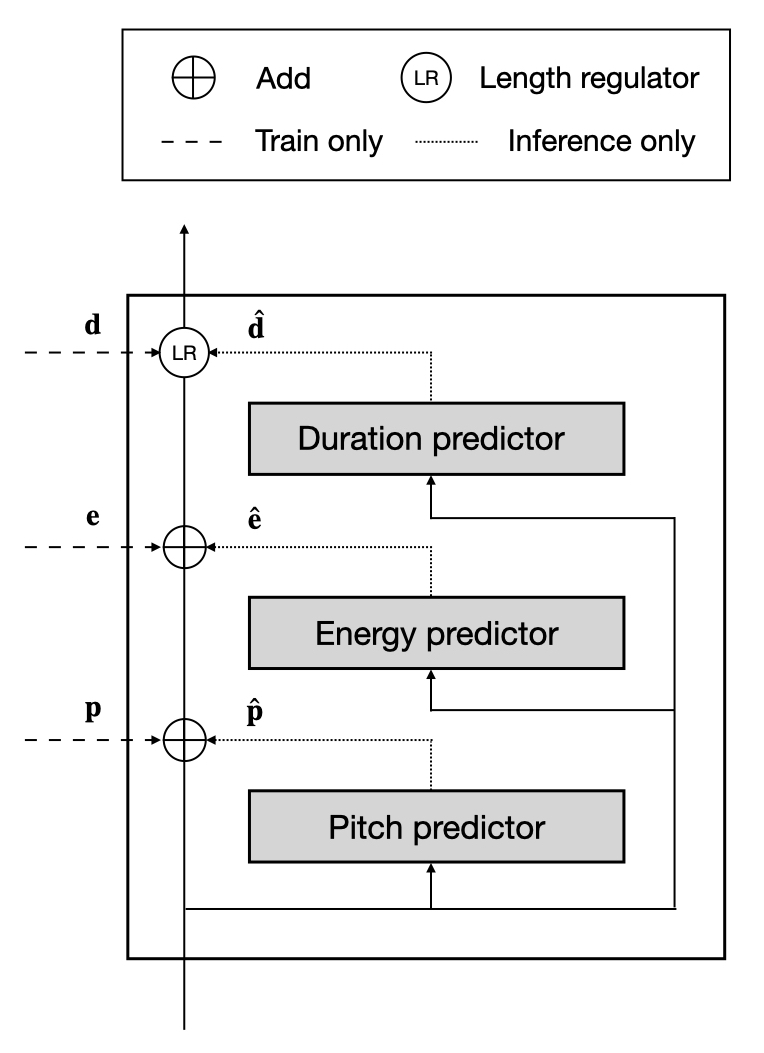}
\caption{Variance adaptor}
\label{fig:var_adaptor}
\end{figure}

\subsection{FastSpeech2}
We adopt FastSpeech2 \cite{DBLP:conf/iclr/0006H0QZZL21} as one of the components of the proposed model. It is a non-autoregressive acoustic feature generator with fast and high-quality speech synthesis. By explicitly modeling token duration with a duration predictor, it improves robustness on synthesis errors such as phoneme repeat and skips. Compared to its previous work; FastSpeech \cite{ren2019fastspeech}, it achieves significant improvement in speech quality by employing additional variance information which is pitch and energy. For our proposed model, We follow the structure of \cite{DBLP:conf/iclr/0006H0QZZL21}, which is a feed-forward Transformer-based \cite{NIPS2017_3f5ee243} encoder, decoder, and 1D convolution-based variance adaptor. Figure \ref{fig:jets} depicts each module in the proposed model. Specifically, the encoder encodes an input text as text embeddings $\mathbf{h}$, and the variance adaptor adds variance information to the text embeddings and expands according to each token duration for the decoder. 

Figure \ref{fig:var_adaptor} depicts the structure of the variance adaptor which consists of pitch, energy, and duration predictor. Pitch and energy predictors are trained to minimize token-wise pitch and energy respectively following the FastSpeech2 implementation of ESPNet2-TTS \cite{hayashi2021espnet2} or FastPitch \cite{lancucki2021fastpitch} instead of frame-wise as in \cite{DBLP:conf/iclr/0006H0QZZL21}. During training, required token-wise pitch and energy $\mathbf{p}, \mathbf{e}$ is computed on the fly by averaging frame-wise ground-truth pitch and energy according to token duration $\mathbf{d}$. The token duration is defined as the number of mel-frame assigned to each input text token and is obtained from the alignment module which will be explained later. After text embeddings are added with pitch and energy, it is expanded by a length regulator (LR) according to the token duration. We use gaussian upsampling with fixed temperature, also known as softmax-based aligner \cite{donahue2021endtoend}, instead of vanilla upsampling by repetition \cite{ren2019fastspeech}.

Note that although we adopt FastSpeech2 for our joint training, we exclude its mel-spectrogram loss so that the proposed model is trained to synthesize raw waveform directly from an input text without intermediate mel-spectrogram. Thus there remains a variance loss that minimizes each variance with $L_2$ loss.

\begin{equation}
    L_{var} = ||\mathbf{d}-\hat{\mathbf{d}}||_2 + ||\mathbf{p}-\hat{\mathbf{p}}||_2 + ||\mathbf{e}-\hat{\mathbf{e}}||_2
\end{equation}
where $\mathbf{d}, \mathbf{p}, \mathbf{e}$ are ground-truth duration, pitch and energy feature sequences respectively whereas $\hat{\mathbf{d}}, \hat{\mathbf{p}}, \hat{\mathbf{e}}$ are predicted ones from the model respectively. 

\subsection{HiFi-GAN}
HiFi-GAN \cite{kong2020hifi} is one of the most famous, GAN-based neural vocoders with fast and efficient parallel synthesis. In the GAN training framework, a model is trained by adversarial feedback where a generator is trained to fake a discriminator, and a discriminator is trained to discriminate between the ground-truth sample and the predicted sample of the generator alternately. Discriminators of HiFi-GAN are designed to improve fidelity by considering a property of speech waveform, which are multi period discriminator (MPD) and multi scale discriminator (MSD). MPD handles diverse periodic patterns of speech waveform whereas MSD operates on the consecutive waveform at different scales with a wide receptive field.

As depicted in figure \ref{fig:jets}, we adopt the HiFi-GAN generator for synthesizing raw waveform from the output of the decoder. HiFi-GAN generator upsamples the output of the decoder through transposed convolution to match the length of the raw waveform where an output of the decoder has the same length as mel-spectrogram of the ground-truth waveform. It has not only adversarial loss but also auxiliary losses which are feature matching loss \cite{kumar2019melgan} and mel-spectrogram loss for the improvement of speech quality and training stability. Note that auxiliary mel-spectrogram loss here is $L_1$ loss between mel-spectrogram of synthesized waveform and that of the ground-truth waveform, which is devised and used for training HiFi-GAN \cite{kong2020hifi}. The auxiliary mel-spectrogram loss is different from the mel-spectrogram loss of FastSpeech2 \cite{DBLP:conf/iclr/0006H0QZZL21}. The training objective of HiFi-GAN follows LSGAN \cite{8237566} and the generator loss consists of an adversarial loss and auxiliary losses as follows.

\begin{equation}
    L_g = L_{g,adv} + \lambda_{fm}L_{fm} + \lambda_{mel}L_{mel}
\end{equation}
where $L_{g,adv}$ is adversarial loss based on least-squares loss function and $\lambda_{fm}, \lambda_{mel}$ is scaling factor for auxiliary feature matching and mel-spectrogram loss respectively.

\subsection{Alignment Learning Framework}
Speech-text alignment is crucial in duration informed networks \cite{ren2019fastspeech}, \cite{yu2020durianx}, \cite{DBLP:conf/iclr/0006H0QZZL21} where the TTS model has a separate duration model and requires explicit duration for its model training as in FastSpeech2. In our proposed model, each token duration $\mathbf{d}$ is used for training a duration predictor, for computing token-averaged pitch, energy from frame-wised ones, and for upsampling the text embeddings. The token duration can be obtained from a pre-trained autoregressive TTS model \cite{li2019neural} as in \cite{ren2019fastspeech} or from speech-text alignment tool such as montreal forced aligner (MFA) as in \cite{yu2020durianx}, \cite{DBLP:conf/iclr/0006H0QZZL21}. Moreover, the training pipeline can be more simplified by incorporating alignment learning so that the required token duration is obtained during the model training on the fly \cite{miao2021efficienttts}, \cite{9747707}, \cite{lim20_interspeech}, \cite{NEURIPS2020_5c3b99e8}.

In this work, we incorporate an alignment learning framework \cite{9747707} into our joint training framework for obtaining the required token duration $\mathbf{d}$ during the training on the fly. An alignment learning framework has shown an improved speech quality as well as fast alignment convergence by devising an alignment learning objective, which can be applied to both autoregressive and non-autoregressive TTS models. An alignment learning objective can be computed efficiently using a forward-sum algorithm. An alignment module in figure \ref{fig:jets} represents the proposed module of an alignment learning framework \cite{9747707}, from which an alignment learning objective as well as each token duration are obtained.

Specifically, an alignment module encodes the text embeddings $\mathbf{h}$ and mel-spectrogram $\mathbf{m}$ as $\mathbf{h}^{enc}, \mathbf{m}^{enc}$ with 2 and 3 1D convolution layers respectively. After that, it computes soft alignment distribution $\mathcal{A}_{soft}$ which is softmax normalized across text domain based on the learned pairwise affinity between all text tokens and mel-frames. 

\begin{align}
    &D_{i,j}  = dist_{L2}(\mathbf{h}^{enc}_{i}, \mathbf{m}^{enc}_{j}) \\
    &\mathcal{A}_{soft} = \texttt{softmax}(-D,dim=0)
\end{align}
where $\mathbf{h}^{enc}_i, \mathbf{m}^{enc}_j$ is the encoded text embeddings and mel-spectrogram at timestep $i, j$ respectively.

From soft alignment distribution $\mathcal{A}_{soft}$, we can compute the likelihood of all valid monotonic alignments which is the alignment learning objective to be maximized.

\begin{equation}
    P(S(\mathbf{h})|\mathbf{m}) = \sum_{\mathbf{s} \in S(\mathbf{h})} \prod^T_{t=1} P(s_t|m_t)
\end{equation}
where $\mathbf{s}$ is a specific alignment between a text and mel-spectrogram (e.g., $s_1=h_1, s_2=h_2, ..., s_T=h_N$), $S(\mathbf{h})$ is the set of all valid monotonic alignments and $T,N$ is the length of mel-spectrogram and text token respectively. A forward-sum algorithm is used for computing the alignment learning objective and we define negative of it as forward sum loss $L_{forward\_sum}$. Notably it can be efficiently trained with off-the-shelf CTC \cite{10.1145/1143844.1143891} loss implementation.

To obtain token duration $\mathbf{d}$, the monotonic alignment search (MAS) \cite{NEURIPS2020_5c3b99e8} is used to convert soft alignment $\mathcal{A}_{soft}$ to monotonic, binarized hard alignment $\mathcal{A}_{hard}$ wherein $\sum^T_{j=1} \mathcal{A}_{hard,i,j}$ represents each token duration. Thus each token duration is the number of mel-frames assigned to each input text tokens and the sum of duration equals the length of mel-spectrogram. There are additional binarization loss $L_{bin}$ which enforces $\mathcal{A}_{soft}$ matches $\mathcal{A}_{hard}$ by minimizing their KL-divergence. Note that we also apply beta-binomial alignment prior as in \cite{9747707}, \cite{shih2021radtts} which multiplies 2d static prior to $\mathcal{A}_{soft}$ to accelerate the alignment learning by making the near-diagonal path more probable.

\begin{align}
    &L_{bin} = -\mathcal{A}_{hard} \odot \text{log} \mathcal{A}_{soft} \\
    &L_{align} = L_{forward\_sum} + L_{bin}
\end{align}
where $\odot$ is Hadamard product and $L_{align}$ is final loss for alignments.

\subsection{Final Loss}
As depicted in figure \ref{fig:jets}, the proposed model consists of the encoder, variance adaptor, decoder, HiFi-GAN generator, and alignment module where the alignment module is used for training only. It is trained to directly synthesize raw waveform from an input text without intermediate mel-spectrogram loss in the GAN training framework. Note that we use discriminators of HiFi-GAN for the training of the proposed model though it is omitted from figure \ref{fig:jets}. Consequently, the loss of the proposed model is GAN training loss integrated with the variance loss and the alignment loss as follows.

\begin{equation}
    L = L_g + \lambda_{var} L_{var} + \lambda_{align} L_{align}
\end{equation}
where we used 1 for $\lambda_{var}$ and 2 for $\lambda_{align}$ as scaling factor of the variance and alignments loss respectively.

\section{Experiments}
For reproducible research, we conducted all experiments including data preparation, model training, and evaluation using ESPNet2-TTS \cite{hayashi2021espnet2} toolkit. The ESPNet2-TTS is a famous, open-sourced speech processing toolkit and it provides various recipes for reproducing state-of-the-art TTS results.

\subsection{Dataset}
We experimented with LJSpeech corpus \cite{ljspeech17} which is an English single female speaker dataset. It consists of 24 hours of speech recorded with a 22.05kHz sampling rate and 16bits. Following the recipe in \texttt{egs2/ljspeech/tts1} in the toolkit, we used 12,600 utterances for training, 250 for validation and 250 for evaluation.

Mel-spectrogram, which is used as an auxiliary loss and an input for an alignment module in the proposed model, was computed with 80 dimensions, 1024 fft size, and 256 hop size. For a fair comparison, g2p-en \footnote{https://github.com/Kyubyong/g2p} without word separators was used as a G2P function, which is the same configuration as the baseline models of ESPNet2-TTS which will be explained later.

\subsection{Model configuration}
We implemented the proposed model using the ESPNet2-TTS toolkit following the configurations and training methods of \texttt{train\_joint\_conformer\_fastspeech2\_hifigan} in the same recipe of the toolkit used for data preparation. The differences are that the transformer was used for an encoder and decoder type instead of a conformer. And we used 256 for the attention dimension and 1024 for the number of the encoder and decoder ff units respectively. In the case of an alignment module, we simply followed the proposed structure in \cite{9747707}. Note that generally a neural vocoder is trained to generate only part of the speech waveform from the corresponding portion of an input sequence for training efficiency. A related hyper-parameter in the toolkit is called segment size which determines the length of the randomly sliced output sequence of the decoder. We used 64 for this hyper-parameter.

For the comparative experiment, we prepared a conventional two-stage, cascaded TTS model as well as another E2E-TTS model. Specifically, we compared the proposed model with state-of-the-art implementations of ESPNet2-TTS. It provides the pre-trained models for public use including CF2 (+joint-ft), CF2 (+joint-tr), and VITS. CF2 (+joint-ft) is Conformer-based \cite{gulati20_interspeech} FastSpeech2 with HiFi-GAN vocoder which are separately trained and jointly fine-tuned. CF2 (+joint-tr) is also Conformer-based FastSpeech2 with HiFi-GAN but it is jointly trained from scratch. VITS is E2E-TTS implementation of the paper \cite{pmlr-v139-kim21f}.

\subsection{Evaluation}
We evaluated the performance of TTS models in objective and subjective metrics. For objective evaluations, mel-cepstral distortion (MCD), log-$F_0$ root mean square error ($F_0$ RMSE), and character error rate (CER) were computed using evaluation scripts provided by the ESPNet2-TTS toolkit. We computed CER using the same pre-trained ESPNet2-ASR model \footnote{https://zenodo.org/record/4030677} which was used in \cite{hayashi2021espnet2}. For subjective evaluation, we conducted a crowdsourced Mean Opinion Score (MOS) test via Amazon Mechanical Turk where each participant, located in the United States, scored each audio sample from different models (including ground-truth audio sample) for naturalness on 5 point scale: 5 for excellent, 4 for good, 3 for fair, 2 for poor, and 1 for bad. Randomly selected 20 utterances from the evaluation set were used for the MOS test and each utterance was listened to by 20 different participants. Audio samples are available online \footnote{https://imdanboy.github.io/interspeech2022}

Table \ref{tab:mos_ljspeech} shows the results on GT (ground-truth recordings), baseline models, and the proposed model. We obtained consistent outcomes as the previous work \cite{hayashi2021espnet2} in that the baseline models achieved high MOS values in the order of CF2 (+joint-ft), CF2 (+joint-tr), and VITS. Interestingly, our proposed model outperformed all of the baselines on MOS as well as objective metrics; $F_0$ RMSE, CER. 

When it comes to the acoustic feature mismatch, the proposed model addressed the problem by the E2E approach which trains the model to generate raw waveform directly from an input text without an intermediate mel-spectrogram. Whereas CF2 (+joint-ft) and CF2 (+joint-tr) solved the problem by jointly fine-tuning and jointly training from scratch respectively. Thus we conjecture that the E2E approach was more effective for improvement than joint fine-tuning or simply joint training of an acoustic feature generator with a vocoder. Another difference compared to CF2 (+joint-ft) and CF2 (+joint-tr) was that the proposed model incorporates alignment learning in its joint training framework. It seems that those factors not only have simplified the training pipeline but also may improve the synthesized speech quality although we didn't investigate how they are related to the model performance thoroughly in this paper.

In the case of VITS, which is also an E2E model with alignment learning capability, it achieved the worst results in our experiment. One of the reasons other than the weakness on the g2p error as reported in \cite{hayashi2021espnet2}, could be its training difficulty due to the somewhat complicated model structure compared to our proposed model. Note that VITS utilizes variational autoencoder and normalizing flow \cite{pmlr-v139-kim21f}.

\begin{table}[t]
  \caption{Results on LJSpeech corpus, where "STD" represents standard deviation and "CI" represents 95\% confidence intervals.}
  \label{tab:mos_ljspeech}
  \centering
  \resizebox{\columnwidth}{!}{%
  \begin{tabular}{lcccc}
    \toprule
    Method              & MCD $\pm$ STD                     & $F_0$ RMSE $\pm$ STD                & CER          & MOS $\pm$ CI                      \\
    \midrule
    GT                  & N/A                               & N/A                                 & 1.0          & 4.08 $\pm$ 0.07                   \\
    \midrule
    CF2 (+joint-ft)     & \textbf{6.73} $\pm$ \textbf{0.62} & 0.219 $\pm$ 0.034                   & 1.5          & 3.96 $\pm$ 0.08                   \\
    CF2 (+joint-tr)     & 6.80 $\pm$ 0.54                   & 0.218 $\pm$ 0.035                   & 1.5          & 3.93 $\pm$ 0.08                   \\
    VITS                & 6.99 $\pm$ 0.63                   & 0.234 $\pm$ 0.037                   & 3.6          & 3.82 $\pm$ 0.09                   \\
    Proposed model      & 7.16 $\pm$ 0.55                   & \textbf{0.215} $\pm$ \textbf{0.034} & \textbf{1.3} & \textbf{4.02} $\pm$ \textbf{0.07} \\
    \bottomrule
  \end{tabular}}
\end{table}

\section{Conclusions}
In this paper, we proposed an end-to-end text-to-speech model which is the jointly trained FastSpeech2 and HiFi-GAN with an alignment module. The proposed model directly generates speech waveform from an input text without intermediate conversion to an explicit human-designed acoustic feature. The training of the proposed model does not have fine-tuning which is required in the two-stage, separately learned text-to-speech models due to the problem of an acoustic feature mismatch. Moreover, we adopt an alignment learning framework so that the proposed model does not depend on external alignment tools for training. Consequently, the proposed model has a simplified training pipeline that is jointly trained in a single stage. For evaluations, we compared the proposed model with publicly available implementations of the ESPNet2-TTS toolkit on the English LJSpeech corpus, and the proposed model achieved state-of-the-art results. It would be interesting for future works to investigate other combinations of joint training other than FastSpeech2 and HiFi-GAN, or to evaluate on multi-speaker dataset.

\bibliographystyle{IEEEtran}

\bibliography{main}


\end{document}